# Dispersive, superfluid-like shock waves in nonlinear optics

Wenjie Wan, Shu Jia, and Jason W. Fleischer
Department of Electrical Engineering
Princeton University

**Abstract**

In most classical fluids, shock waves are strongly dissipative, their energy being quickly lost through viscous damping. But in systems such as cold plasmas, superfluids and Bose–Einstein condensates, where viscosity is negligible or non-existent, a fundamentally different type of shock wave can emerge whose behaviour is dominated by dispersion rather than dissipation. Dispersive shock waves are difficult to study experimentally, and analytical solutions to the equations that govern them have only been found in one dimension (1D). By exploiting a well-known, but little appreciated, correspondence between the behaviour of superfluids and nonlinear optical materials, we demonstrate an all-optical experimental platform for studying the dynamics of dispersive shock waves. This enables us to observe the propagation and nonlinear response of dispersive shock waves, including the interaction of colliding shock waves, in 1D and 2D. Our system offers a versatile and more accessible means for exploring superfluid-like and related dispersive phenomena.

Unlike dissipative shock waves in ordinary gases/fluids, which have a well-defined shock front due to viscosity, dispersive superfluid-like shock waves have an oscillatory front. These oscillations result from two basic, and related, properties of the superfluid state: nonlinearity and coherence. Coherence results from cooling the fluid, so that the constituent particles of the condensate are perfectly correlated, while nonlinearity refers to the inter-particle interactions which make this correlation possible. For different reasons, these two properties also appear in nonlinear optics. While the relationship is well known in condensate community [e.g. nonlinear "atom optics" studies in Bose-Einstein condensates (BEC)[1-3]], the relationship has been underappreciated from the opposite perspective. Here, we build on previous theoretical[4, 5] and experimental[6, 7] work on superfluid behavior in BEC to examine the optical equivalent of condensate shock waves. We demonstrate basic dispersive, dissipationless shock waves in one and two transverse dimensions, characterize their nonlinear properties, and reveal the nontrivial interactions when two such shocks collide.

While dispersive shock waves in optics have been studied previously for temporal pulses in fibers[8-16], they have not yet been considered in the spatial domain. In this case, the extra dimensional freedom allows consideration of wavefront geometry, which is shown to significantly affect shock propagation and interaction. The particular system considered here is a spatial one in which a continuous optical wave propagates in a nonlinear Kerr-like medium, mainly along the z-axis. To an excellent approximation, the slowly-varying amplitude $\psi$ of such a field can be described by the nonlinear Schrödinger equation:

$$i\frac{\partial \psi}{\partial z} + \frac{1}{2k_0}\nabla_\perp^2 \psi + \Delta n(|\psi|^2)\psi = 0 \qquad (1)$$

where $k_0 = 2\pi n_0/\lambda$ is the wavenumber, $\lambda/n_0$ is the wavelength in a homogeneous medium of refractive index $n_0$, and $\Delta n = n_2 k_0 |\psi|^2/n_0$ is the nonlinear index change for a Kerr medium with coefficient $n_2$ ($n_2 < 0$ for defocusing). For the spatial case, the transverse Laplacian describes beam diffraction, while in the temporal case it describes pulse spreading due to dispersion. As is well-known[17-20], Eq. (1) also describes the (macroscopic) ground-state wavefunction for a fully-condensed quantum state: $i\hbar \partial_t \psi + (\hbar^2/2m)\nabla_\perp^2 \psi + g|\psi|^2 \psi = 0$, where $m$ is the mass of the underlying particle and the nonlinear coefficient $g$ represents the mean-field contribution of (s-wave) interactions. In this approximation, the dynamics are more properly described as wave mechanical rather than quantum, with $\hbar$ simply serving as a parameter which normalizes the wavefunction. Note in particular that wavepacket evolution in time corresponds to beam propagation in space.

It is instructive to give the fluid context for the dynamics of Eq. (1). Applying the Madelung transformation[21] $\psi(x,z) = \sqrt{\rho(x,z)}\exp[iS(x,z)]$, where $\rho$ is the intensity of the beam and $S$ is its coherent phase, and scaling $(x,z) \rightarrow k_0(x,z)$ gives the Euler-like fluid equations[22, 23]:

$$\frac{\partial \rho}{\partial z} + \nabla_\perp \cdot (\rho v) = 0 \tag{2}$$

$$\frac{\partial S}{\partial z} + \frac{1}{2}v^2 + c^2\left(\frac{\rho}{\rho_\infty} - 1\right) - \frac{1}{2}\left(\frac{1}{\sqrt{\rho}}\nabla_\perp^2 \sqrt{\rho}\right) = 0 \tag{3}$$

Here, $v = \nabla_\perp S$ is the "fluid" velocity and $c = \sqrt{n_2 \rho_\infty / n_0}$ is an effective "sound" speed due to the background intensity $\rho_\infty = \rho(|z| \rightarrow \infty)$. The last term in Eq. (3), often called the "quantum pressure" in condensed matter, is significant only for steep gradients and in regions where the fluid density/optical intensity goes to zero, e.g. wave-breaking, dark soliton formation, and the "healing" of a condensate near a boundary[23].

The experiments below consider a bright hump superimposed on a uniform, low-intensity background (Figs. 1 and 2). In the initial stages of evolution, the last term in Eq. (3) can be neglected, giving the standard momentum equation

$$\rho\left(\frac{\partial v}{\partial z} + v\nabla_\perp v\right) = -\nabla_\perp\left(\frac{n_2 \rho^2}{2n_0}\right), \tag{4}$$

In this form, it is clear that the nonlinearity gives rise to an effective pressure, whose gradient drives the acceleration of the optical fluid. Note from Eq. (1) that the nonlinear contribution to the phase $S \sim n_2 k_0 |\psi|^2 (\Delta z)/n_0 = n_2 k_0 \rho (\Delta z)/n_0$, so that Eq. (4) is self-consistent with the definition of velocity.

The dynamics of an initial profile depend on the strength of the nonlinearity. For concreteness, consider a 1D Gaussian intensity profile superimposed on a homogeneous background: $\rho(x,0) = \rho_\infty + 2\eta\rho_\infty \exp(-x^2/\sigma^2)$. In the linear case, equivalent to a non-interacting gas, the hump will simply diffract (disperse) against the background. In the nonlinear case, the hump will split into two equal pieces which repel each other. For weak nonlinearity, the basic physics can be seen by considering small perturbations around the background intensity, i.e., $\eta \ll 1$. In this case[4], Eqs. (2) and (4) give a sound-like propagation equation, $(\partial_z^2 - c^2 \partial_x^2)S = 0$, resulting in two traveling waves: $\rho(x,z) = \rho_\infty + \eta\rho_\infty \left[\exp(-(x-cz)^2/\sigma^2) + \exp(-(x+cz)^2/\sigma^2)\right]$. For stronger nonlinearity, the two pieces will propagate with a velocity $v = v(\rho)$ which depends on the local intensity, rather than at the constant sound speed *c*. Higher-intensity parts of the profile will travel at faster speeds, leading to wave steepening and eventual shock formation[24]. Note that both the initial hump splitting and shock formation require a background intensity; without a reference (even in the nonlinear case), there is only a single hump which weakens as an expanding rarefaction wave.

Unlike shock models with viscosity (e.g. Burgers-type descriptions[25, 26]), there is no dissipation mechanism to counteract nonlinear wave steepening. Instead, the increasing gradient triggers an increase in dispersion. More accurately, self-phase modulation within the high-intensity region generates new (spatial) frequencies, which then disperse into the surrounding medium. Rather than a well-defined front in which the high pressure/intensity monotonically decreases to match the low-pressure background, the traveling wave develops on oscillating front (Fig. 2c). Here, the presence of a background provides a reference intensity/phase for visualizing the different wave components. In condensate terms, the background density sets the reference sound speed, meaning that higher perturbation densities naturally correspond to supersonic sources. As shown by several authors[5, 27-29], the 1D shock profile is a Jacobi elliptic function, found by matching the high- and low-intensity boundaries. The inner, nonlinear part of the front resembles a train of dark (or gray) solitons, while the outer part is a low-intensity region with oscillations that are effectively sound-like[5, 27-31].

In most systems in which shock waves are possible, there is usually a thermal component responsible for energy dissipation. Moreover, viscosity usually dominates dispersion in cases where both elements are present, damping oscillations before they can form. Here, the focus is on the basic nonlinear dynamics of dispersive waves, without the complications of a viscous term. In the ideal case, the model system is a fully-condensed superfluid, in which excitations are ignored (so that the mean-field approximation of Eq. (1) is valid[23]). Similarly, cold plasmas can support such dispersive waves when damping effects can be neglected[32-35]. In hydraulics, neglecting viscosity gives an "inertia"-dominated regime, suitable for coherent descriptions of dam breaking, surface waves, and undular bore propagation[24]. (An alternative view of this can be obtained by considering Eq. (1) as the long-wavelength limit of other dispersive wave models, e.g. the Korteweg-de Vries equation[36].) From an experimental viewpoint, the mapping to nonlinear optics allows the isolation of a totally-coherent wave, so that the basic properties of shocks in a purely dispersive fluid can be studied in detail. It also greatly simplifies the setup (shown in Fig. 1), and provides easy control of the input conditions and direct imaging of the output.

Experiments were performed using 532nm laser light projected into an 8x8x8mm SBN:75 photorefractive crystal. For this crystal, the nonlinear index change in Eq. (1) is $\Delta n = -(1/2) n_0^3 r_{33} E_{app} \bar{\rho}/(1+\bar{\rho})$, where $n_0 = 2.3$ is the base index of refraction, $r_{33} = 1340$ *pm/V* is the appropriate electro-optic coefficient with respect to the applied field $E_{app}$ and the crystalline axes, and the relative intensity $\bar{\rho}$ is the input intensity $|\psi|^2$ measured relative to a background (dark current) intensity[37, 38]. A self-defocusing nonlinearity is created by applying a voltage bias of -500V across the crystal and taking advantage of the photorefractive screening effect. *This voltage is held constant throughout the nonlinear experiments, and only the intensity of the central hump is changed to probe nonlinearity.* This restriction isolates the dynamics to intensity-dependent effects only, ensuring the generality of the results. Note further that the use of defocusing nonlinearity minimizes the difference between the saturable and Kerr cases[39, 40]. For the shock waves considered here, there is less evolution (higher central intensity and fewer front oscillations) in a given length for saturable nonlinearities than for the Kerr case; otherwise, the two behaviors are identical.

The experimental setup is shown in Fig. 1. Extraordinarily-polarized laser light is split using a Mach-Zehnder interferometer: a weak plane wave in one of the arms serves as the low-intensity background, while the central intensity hump is formed by using a lens (cylindrical or spherical) in the other arms. The beams are then recombined on the input face of the crystal. The language used here is "hump-on-background," but it is important to emphasize that, as far as the crystal is concerned, the input wavefunction is a single coherent wave. At the exit face of the crystal, the output beam profile is imaged onto a CCD camera. Real-space imaging allows photographs of position (**x**) space, while performing an optical Fourier transform allows photographs of momentum (**k**) space.

Typical shock waves are presented in Fig 2. Initial stripe, elliptical, and circular profiles (Fig. 2, top row) were formed by using cylindrical lenses (one for the stripe, two orthogonal ones for the ellipse) and one circular lens, respectively. The intensity ratio between peak and background was adjusted by a variable attenuator placed before the lenses. For the inputs here, the background beam has 10mW of power and the peak-to-background ratio is 20:1. In the linear case (Fig. 2, middle row), the high-intensity humps simply diffract against the low-intensity background, keeping their Gaussian-like structure and creating small ripples in the tails as the phase front curves. In contrast, turning on the defocusing nonlinearity (Fig. 2, bottom row) forces the hump apart, depleting the central region and creating two repulsive shock waves with oscillating fronts. Note that the stripe and circle profiles are symmetrical, while the 3:1 ellipse has an asymmetric profile, as expected from the anisotropic intensity gradient.

As the intensity ratio of the initial profile increases, the shocks become more violent, with faster wave propagation and more oscillations within the front (stronger effective repulsion and higher nonlinear phase). Indeed, dimensional analysis from Eq. (3) suggests that $v \sim \sqrt{\rho/\rho_\infty}$.

However, simple scaling arguments cannot determine the coefficient. In Fig. 3, we plot the measured front length (measured from the centerline to the end of oscillations) as a function of $\rho/\rho_\infty$. The solid curves are best fits of the functions $D_s = a_s(1 + b_s\sqrt{\rho/\rho_\infty})$, done independently for each shape *s*. Here, $a_s = L\sqrt{n_0^3 r_{33} E_{app} \rho_\infty / 2} = 54 \mu m$ is a dimensional scaling constant, dependent only on the background intensity and fixed crystal properties (length *L*, base index $n_0$, electro-optic coefficient $r_{33}$, and applied voltage $E_{app}$), and the *b*-coefficients are 1.2, 1.0, 0.92 ± 0.04 for the stripe, ellipse, and circle, respectively. The power scaling matches the predicted relation, but the stripe coefficient is higher than recent 1D theory[4, 5, 30, 31] suggesting *b* = 1.0 (probably due to our use of very high intensity ratios). Note, however, that there has been *no* analytic treatment of dispersive, dissipationless shock waves in higher dimensions. The experimental results here show that geometry and the available expansion directions play a significant role.

Further insight into the behavior of dispersive shock waves can be gained by considering their basic interactions. In Figs. 4-6, several types of shock collision are shown. For these experiments, a second lens arm is added to the Mach-Zehnder scheme shown in Fig. 1. Fig. 4 shows typical results from 1D shock interactions. Figs. 4a-c show, respectively, output profiles when the initial humps are separated by 500, 200, and 50μm. In Fig. 4a, the shocks do not intersect over the crystal length (and therefore show individual profiles), while in Fig. 4b the initial condition is chosen so that the waves do not intersect in the linear case but do overlap in the nonlinear case. Despite the low intensity in the leading edges, the profile shows that shock collision is an inherently nonlinear process. As shown in Fig. 4b, the collision region has 1) a lower maximum intensity than the expected 4x gain of linear superposition, 2) an internal period of 7μm, significantly more than the 5μm expected from a linear sum of 10μm tails, 3) a narrower width than that of the individual fronts, and 4) a more regular period than the individual tails. The first two characteristics are a direct result of the defocusing (repulsive) nonlinearity, while the last two involve details of nonlinearly interacting waves that are still being explored. Four-wave mixing effects are particularly relevant here[41-43], but the dynamics are complicated by the broad spectrum of spatial scales within the shock fronts.

For closer initial separations, the individual shock profiles cannot form, and a different aspect of the dynamics becomes dominant. As shown in Fig. 4c, the output consists of a single shock with a double front, rather than two individual shocks with a common collision region. Essentially, the initial overlap creates a high-intensity region, which itself acts as a source for a new shock wave. This nonlinear Huygens' (or Hadamard) principle[44, 45] is common to all shock wave interactions. Indeed, we note that similar results were observed in the BEC experiments of Ref. [7]. In that case, however, the presence of a trap potential created transverse variations in the density. The resulting variations in shock speed across the front led to hybrid shock-vortex structures[7]; in interactions, the vortices can split and merge, giving rise to rich and complex dynamics that are coupled with the shock-shock interactions. By contrast, the photonics experiments here focus on the homogeneous case. Remarkably, it is found that the shock fronts are stable during propagation and do not generate vortices even after (head-on) collision. We conjecture that the *array-like* structure of the front is responsible for this, as individual 1D dark solitons suffer a snake instability (leading to vortices) in two transverse dimensions[46, 47] but 1D arrays are stable[48, 49].

Power spectra of 1D shock interactions, obtained by performing on optical Fourier transform on the output profiles in Fig. 4, are shown in Fig. 5. The linear reference case, that of two widely-separated Gaussian beams on a background, is shown in Fig. 5a. There are three main features of this spectrum: 1) there is a dominant central peak at $k_x=0$ due to the uniform background, 2) there is a fast oscillation resulting from the spectral beating $\exp[-(x-\Delta)^2] + \exp[-(x+\Delta)^2] \rightarrow \cos(k\Delta) \exp[-k^2]$, and 3) there is a slow envelope modulation from wave mixing with the central background peak. The equivalent nonlinear case of two widely-separated shock waves (Fig. 4a) is shown in Fig. 5b; as in Fig. 5a, it is a modulated form of the individual power spectrum. By comparison, the shock spectrum consists of a much broader range of spatial frequencies, with two spectral holes appearing within the linear range. These holes create two distinct spectral regions, or humps, on either side of the central peak. The inner regions are large-scale modulations resulting from the initial splitting of the hump, while the outer tail regions result from wave steepening and the nonlinear generation of dispersive waves (much like the broad spectrum in supercontinuum generation[50]). As the initial beams are brought closer together, the fronts will overlap and interact with each other during propagation. In terms of the spectral energy density, there will be a power transfer between the two regions highlighted in Fig. 5b. Difference frequencies in the (small-scale) tails will transfer energy back to the large-scale humps. (Due to the broad background, the power within the central peak stays relatively constant.) As shown in Fig. 5e, there is a maximal amount of (integrated) energy transfer as a function of initial shock separation, occurring at a distance which corresponds with the front width. For closer initial separations, the double-front shock of Fig. 4c is formed; in this case, the tails do not have time to form initially, so the interaction results in energy transfer from the large-scale humps to smaller-scale waves. Finally, we note that the collision dynamics, and the corresponding spectral energy distributions, are very sensitive to the relative phase of the shocks.

In higher-dimensional collisions, wavefront geometry becomes a significant factor. Figure 6 shows experimental results of 2D shock interactions along with simulation results from a split-step Fourier beam propagation code. In the collision of two equal circular shocks (Figs. 6c,d), the ripples penetrating each ring are *straight*, rather than the circles expected from a linear superposition (e.g. drops in a pond). Again, this is due to the nonlinear Huygens' principle: the two intersecting arcs originally superimpose to form a straight front, which then acts as a source for quasi-1D shock waves. Note also that the central "peak" has split into two due to the self-defocusing nonlinearity. Similar wavefront distortions occur in the 1D-on-2D collision (Figs. 6e,f). The right-moving shocks have a weaker curvature than they started with (compare with the undisturbed rings on the left), while the left-moving shocks have a concave front. This type of curvature would normally create a lensing effect, but the defocusing nonlinearity provides a competing force.

In conclusion, we have demonstrated dispersive, dissipationless shock waves in nonlinear optics. Initial conditions consisted of high-intensity humps defocused against a uniform background, while the output profiles consisted of two repulsive waves with oscillating fronts. Compared with the linear case, the nonlinear shock system has a depleted central region and a broader spectral range, implying more efficient energy dispersion. Higher-intensity perturbations result in faster shocks, with a speed that depends on expansion geometry. Interactions between shocks are complex, with nontrivial energy exchanges that depend on details of the collision region. Intuition is helped by invoking a nonlinear Huygens' principle, in which linear superposition of initial waves results in a nonlinear source of new shocks. This is seen most clearly in 2D interactions, where the wavefront geometry is significantly modified by interactions. The internal dynamics are complex, and have not yet been examined in any rigorous detail. While such behavior occurs in any dissipationless, coherent wave system, such as idealized, non-viscous hydrodynamics and fully-condensed systems, observations are significantly easier in the optical case. Hence, in addition to providing a versatile platform for new photonic physics, it is anticipated that the results reported here will lead to all-optical modeling of even richer (super)fluid-like phenomena in the near future.

**ACKNOWLEDGMENTS**

We thank M.P. Haataja, C.B. Arnold, and M. Warnock-Graham for useful discussions. This work was supported by the NSF and AFOSR.

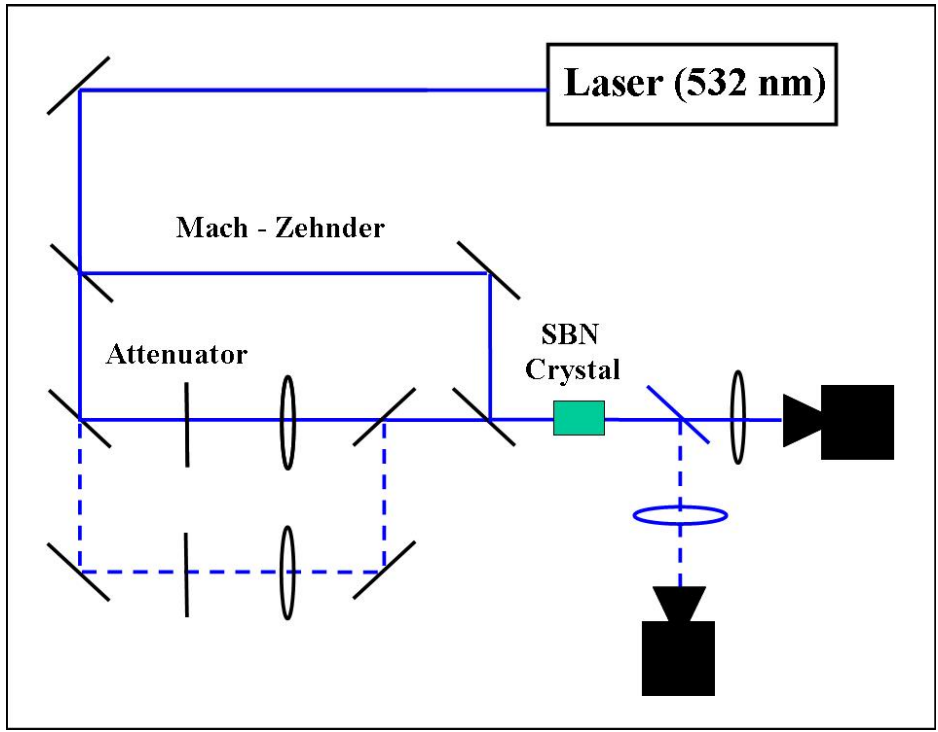

**Fig. 1.** Experimental setup. Light from a laser is split using a Mach-Zehnder interferometer. A cylindrical/circular lens placed in one of the arms focuses a beam onto the input face of an SBN:75 crystal. For the nonlinear experiments, a *constant* voltage of -500V is applied across the crystalline c-axis to set the photorefractive screening effect, while the shock strength is controlled by varying the hump:background intensity ratio with an attenuator. Light exiting the crystal is then imaged into a CCD camera. Both position (**x**) space and momentum (**k**) space are imaged. For the collision experiments, a second lensing arm in the interferometer is added.

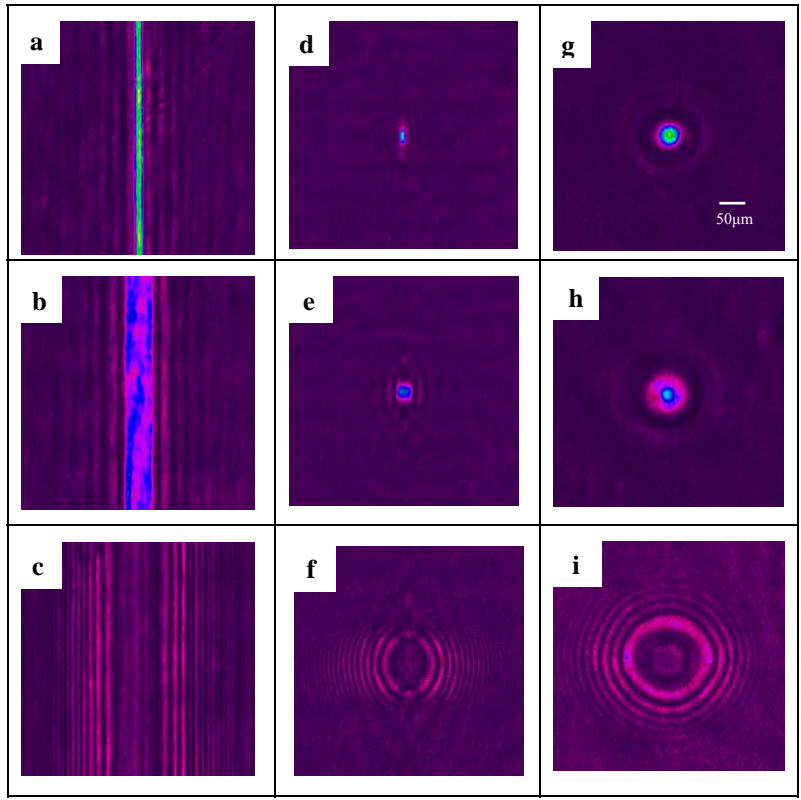

**Fig. 2**. Experimental pictures of superfluid-like optical spatial shock waves. Top row: input face. Middle row: linear diffraction at output face. Bottom row: nonlinear shock waves at output face. (a-c) 1D stripe. (d-f) 2D ellipse. (g-i) 2D circle.

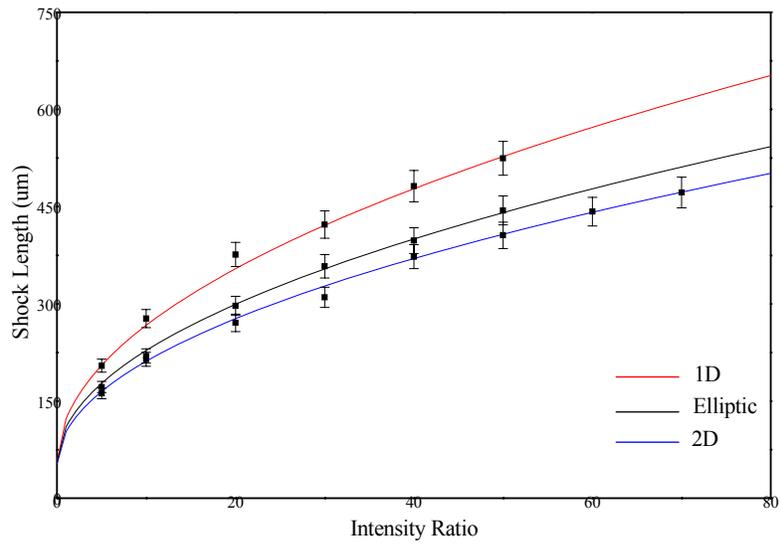

**Fig. 3.** Shock length, measured from the centerline to the end of oscillations, with respect to peak-to-background intensity ratio. The error bars signify maximum measured deviations due to poor visibility of the leading edge endpoint. Top to bottom: three solid curves are plots of the functions $D_s = a_s + b_s \sqrt{\rho/\rho_\infty}$ to fit the 1D stripe, 2D ellipse and 2D circle cases, respectively.

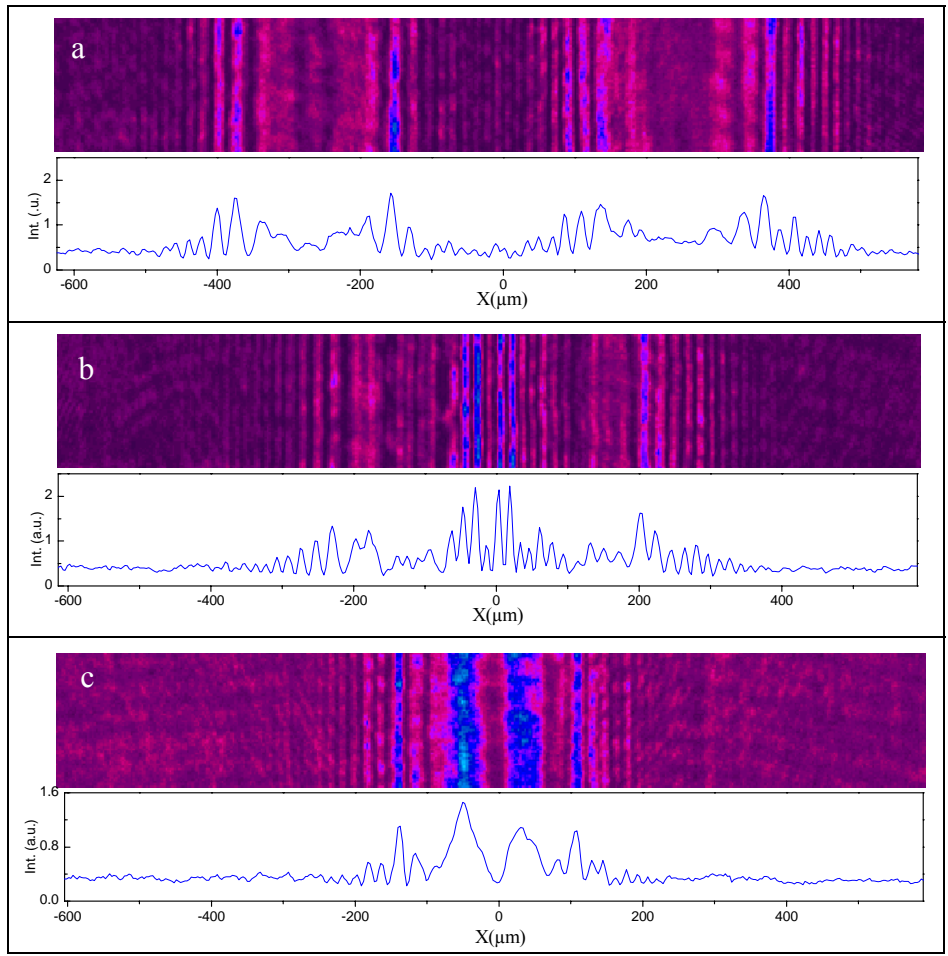

**Fig. 4**. Experimental output pictures vs. initial separation distance between two adjacent shocks. Experimental output pictures and cross-sections when the initial separation distance is: (a) 500μm (b) 200 μm (c) 50 μm. Note that there is no collision in (a) within the propagation distance, (b) shows a typical collision process, and (c) shows a single, double-front shock output due to very close interactions.

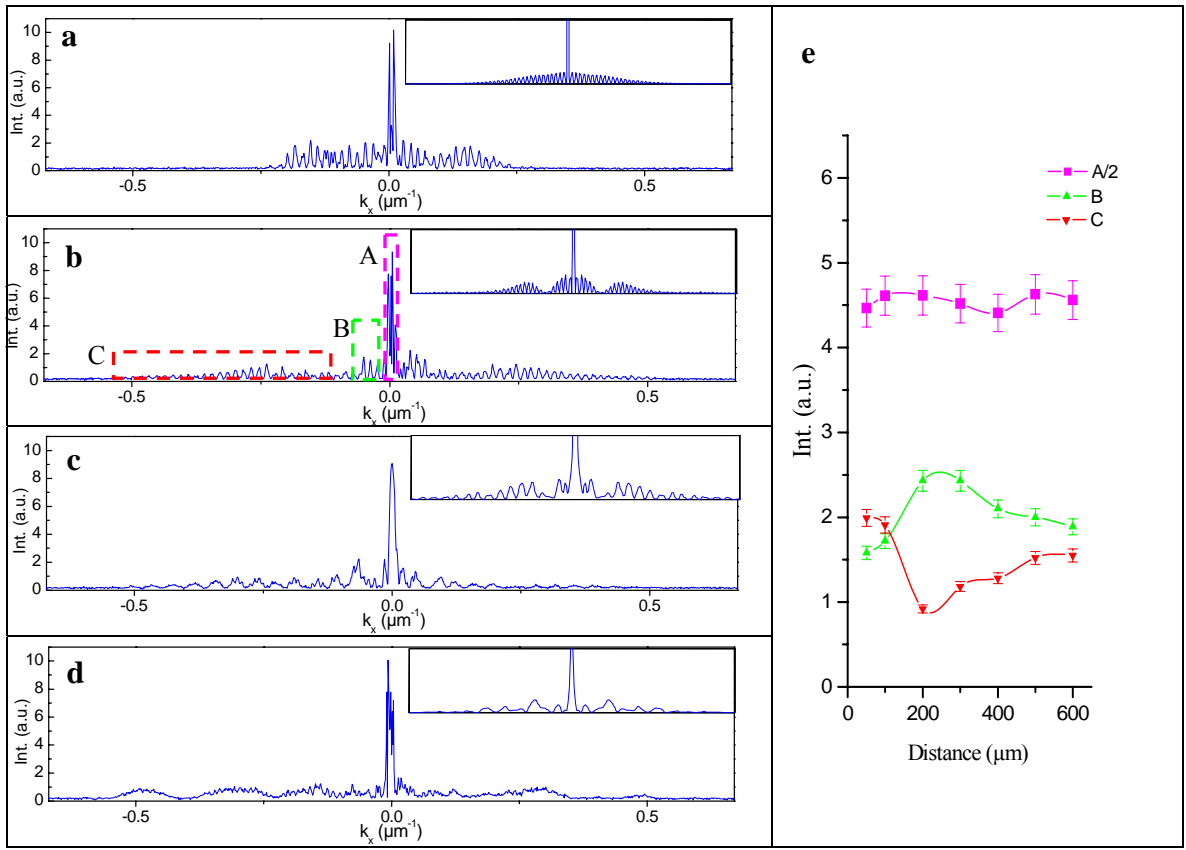

**Fig. 5**. Fourier power spectra of shock collisions vs. initial separation distance. Experimental output measurements of (a) Linear diffraction from an initial hump separation of 500μm (b-d) Nonlinear spectra corresponding to Figs. 4a-c: (b) initial 500μm separation (c) initial 200μm separation (d) initial 50μm separation. Insets are simulation results from beam propagation code. (e) Average value of spectral windows A,B,C vs. initial separation distance. The solid lines in (e) are curves to lead the eye, while the error bars indicate standard deviation of measurements arising from phase fluctuations.

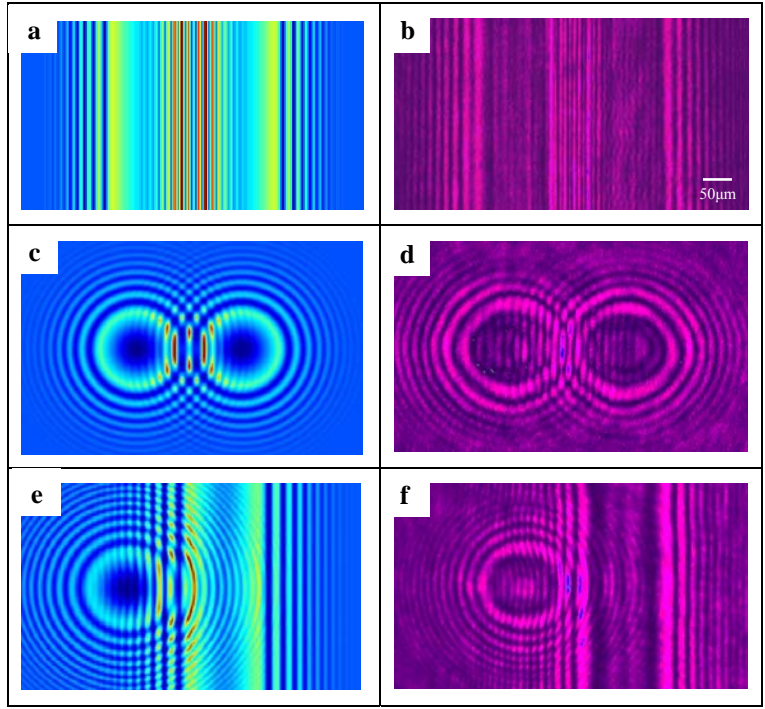

**Fig. 6**. Shock wave collisions. Left column: beam propagation simulations. Right column: experimental output pictures. (a,b) 1D collision. (c,d) 2D collision. (e,f) 1D-on-2D shock collision. The outer regions show undisturbed shock behavior, while the inner regions reveal the nontrivial interaction of nonlinear, dispersive waves. In particular, the wavefronts penetrating each circle in (c,d) are straight, the right-going wavefronts in (e,f) are flattened, and the left-going wavefronts in (e,f) become concave.